\documentstyle{article}
\input{amssym.def}

\begin{document}

\begin{flushleft} \Large
{\bf SUPERSYMMETRY AND SUPERCOHERENT STATES OF A
NONRELATIVISTIC FREE PARTICLE}
\end{flushleft}

\begin{quotation}
{\normalsize \noindent
Boris F. Samsonov
\footnote {Electronic mail address: samsonov@phys.tsu.tomsk.su

J. Math. Phys. {\bf 38} (9), September 1997, p. 4492-4503.
}
\vspace{5mm}
\noindent {\it  Tomsk State University, 36 Lenin Avenue, 634050 Tomsk, Russia}

\vspace{5mm}
\noindent (Received 20 October 1996; accepted for publication 16 May 1997)

\vspace{5mm}
\noindent Coordinate atypical representation of the orthosymplectic superalgebra $%
osp(2/2)$ in a Hilbert superspace of square integrable functions constructed
in a special way is given. The quantum nonrelativistic free particle
Hamiltonian is an element of this superalgebra which turns out to be a
dynamical superalgebra for this system. The supercoherent states, defined by
means of a supergroup displacement operator, are explicitly constructed.
These are the coordinate representation of the known atypical abstract super
group $OSp(2/2)$ coherent states. We interpret obtained results from the
classical mechanics viewpoint as a model of classical particle which is
immovable in the even sector of the phase superspace and is in rectilinear
movement (in the appropriate coordinate system) in its odd sector. \
\copyright \ {\it 1997 American Institute of Physics.} [S0022-2488(97)00809-8]
}
\end{quotation}

\pagenumbering{arabic}

\Roman{section}

\vspace{0.5cm}

\begin{flushleft}
{\large {\bf I. INTRODUCTION} }
\end{flushleft}

\vspace{0.5cm}
The supersymmetry in physics has been introduced in
the quantum field theory for unifying of interactions of different kinds in
a unique construction \cite{r1}. Supersymmetric formulation of quantum
mechanics is due to the problem of spontaneous supersymmetry breaking \cite
{r2}. Ideas of supersymmetry have been profitably applied to many
nonrelativistic quantum mechanical problems since, and now there are no
doubts that the supersymmetric quantum mechanics has its own right to exist
(see for Ref. \cite{Coop} a recent review). It is worth noticing that almost
all papers are concerned with the stationary Schr\"odinger or Pauli
equations. There are only a few papers dealing with nonstationary equations
\cite{BS,KMNT}.

The mathematical foundation of the conventional the quantum
mechanics consists in the operator theory in a Hilbert space \cite
{Neum,Simon}. The notion of Hilbert space is also indispensable in the
construction of unitary representations of Lie groups. The space of square
integrable on Lebesgue measure functions is one of the most important
realization of the Hilbert space.

When we pass from the conventional quantum mechanics to the
supersymmetric ones and from conventional Lie algebras and groups to
superalgebras and supergroups we need the notion of Hilbert superspace.
There are few works about Hilbert superspaces \cite{DW,HSS}. It seems that
mathematically rigorous and consistent theory of Hilbert superspaces and the
theory of operators acting in these superspaces need to be developed.

In this paper for a particular case of nonrelativistic free particle
we construct a Hilbert superspace which is a $z_2$-graded
infinite-dimensional linear space equipped with a super-Hermitian form
(superscalar product) and in some sense complete. Solutions of the free
particle Schr\"odinger equation form a dense set in this superspace.

The notion of coherent states is widely used in the conventional
quantum mechanics and mathematical physics \cite{P}-\cite{MalMan}. Many
definitions of coherent states exist \cite{Lxxv}. The more suitable for
generalization to the supersymmetric case is the one based the on
group-theoretical approach developed by Perelomov for a wide class of
conventional Lie groups \cite{P}. This definition has a natural
generalization to Lie supergroups and superalgebras based on the notion of
supergroup translation operator \cite{FKNT}.

In a number of papers \cite{Ber,K} abstract representations of some
simple Lie superalgebras are studied and with their help supercoherent
states have recently been constructed \cite{KMNT,FKNT},\cite{BSB}-\cite{GN}
and underlying geometric structure has been envisaged \cite{G,GN}.
Nevertheless, application of these results to the quantum mechanics is not
numerous \cite{KMNT}, \cite{FKNT}. The supercoherent states of a charged
spin-$\frac 12$ particle in a constant magnetic field \cite{FKNT} and in a
time-varying electromagnetic field \cite{KMNT} are explored and interpreted
in a physical context. The role of Grassmann variables is clarified and
insight is gained into the link between supercoherent states and the
classical motion. The fermion monopole system which is known to have a
dynamical $OSp(1/2)$ supersymmetry is considered and their supercoherent
states are obtained \cite{FKNT}.

In this paper we show that not only the above-mentioned quantum
systems may be interpreted in terms of supersymmetric notions but every
nonrelativistic one-dimensional quantum system with quadratic in coordinate $%
x$ Hamiltonian exhibits supersymmetric properties. We concentrate our
attention on a simple but nontrivial case of the nonrelativistic free
particle which in our interpretation has $OSp(2/2)$ dynamical supersymmetry.
More precisely, the space of solutions of the Schr\"odinger equation for the
free particle is an atypical Lie $osp(2/2)$-module. We use the notion of
supergroup both as a $z_2$-graded group and a superanalytic supermanifold
\cite{Rog}. The action of a supergroup operator translation is defined on a
dense set in the Hilbert superspace. This operator maps the dense set from
the Hilbert superspace onto the Grassmann envelope of the second kind of the
Hilbert superspace. Being applied to a maximal symmetry vector (in our case
the vacuum vector) this operator produces supercoherent states for the free
particle which are the coordinate representation of the known atypical $%
osp(2/2)$ coherent states. These states are parametrized by the points of
the $N=1$ superunit disk ${\cal D}^{(1|1)}$. The supermanyfold ${\cal D}%
^{(1|1)}$ is a phase superspace of a classical system possessing remarkable
property, namely, geometric quantization of this system gives a
superholomorphic representation of the initial (i.e., free particle) quantum
system. By these means we construct a classical mechanics system which
corresponds to the nonrelativistic quantum free particle. Finally we
interpret the obtained classical system as a classical particle which is
immovable in the even sector of the phase superspace and is in rectilinear
movement in its odd sector.

The paper is organized as follows. In Sec. II we summarize the
well-known results about the representation in the Hilbert space of symmetry
algebra of the free particle Schr\"odinger equation we need further. Section
III includes two parts. In the first one we recall main notions about the
superanalysis and in the second we construct a Hilbert superspace of square
integrable functions. In Sec. IV we define the action of operators in
constructed Hilbert superspace which are symmetry operators for the free
particle Schr\"odinger equation and realize an atypical coordinate
representation of the $osp(2/2)$ superalgebra. In Sec. V the coherent states
for the nonrelativistic free particle are constructed. In Sec. VI we discuss
obtained results, compare them with the known ones, and interpret from the
classical mechanics viewpoint.

\vspace{0.5cm}

\begin{flushleft}
{\large {\bf II. SCHR\"ODINGER ALGEBRA }}
\end{flushleft}

\vspace{0.5cm}
In this section we summarize briefly the well-known
constructions \cite{M} for a representation in the Hilbert space of square
integrable on Lebesgue measure functions of the Schr\"odinger algebra ${\cal %
G}_2$ which is a dynamical symmetry algebra for the nonrelativistic free
particle.

Consider the free particle Schr\"odinger equation
\begin{equation}
\label{e1}i \partial _t\chi (x,t)=h\chi (x,t),\quad h=-\partial _x^2,\quad
\partial _x^2=\partial _x\cdot \partial _x,\quad \partial _x=\partial
/\partial _x.
\end{equation}
Solutions of this equation pertaining to the space $L_2({\Bbb R})$ of square
integrable functions on full real axis with respect to Lebesgue measure are
well known \cite{M}:
\begin{equation}
\label{bf}
\begin{array}{c}
\chi _n(x,t)=\langle x\mid n,t\rangle =\left( -
i \right) ^n\left[ n!\sqrt{2\pi }\left( 1+i t\right) \right] ^{-1/2} \\
\times \exp \left( -
i n\arctan t-\frac{x^2}{4+4i t}\right) He_n(z), \\ z=x/\sqrt{1+t^2},\quad
He_n(z)=2^{-n/2}H_n(z/\sqrt{2}),
\end{array}
\end{equation}
where $H_n(z)$ are the Hermitian polynomials. Let ${\cal L}$ be lineal
(linear hull) of the functions $\left\{ \chi _n(x,t)\right\} $. Introduce
the notations $\psi _n(x,t)=\chi _{2n}(x,t)=\chi _{2n}(-x,t)$ and $\varphi
_n(x,t)=\chi _{2n+1}(x,t)=-\chi _{2n+1}(-x,$ $t)$, $n=0,1,2,\ldots $ and denote
${\cal L}^0$ the lineal of even functions $\left\{ \psi _n(x,t)\right\} $
and ${\cal L}^1$ the lineal of odd ones $\left\{ \varphi _n(x,t)\right\} $.

If we introduce the scalar product (Hermitian form) in ${\cal L}$ in
a usual way,
\begin{equation}
\label{e2}\langle \chi _1(x,t)\mid \chi _2(x,t)\rangle =\int_{-\infty
}^\infty \overline{\chi }_1(x,t)\chi _2(x,t)dx,
\end{equation}
where overline signifies the complex conjugation and the integral should be
understood in the sense of Lebesgue, then the completion $\overline{{\cal L}}
$ of the lineal ${\cal L}$ with respect to the measure induced by this
scalar product gives the Hilbert space $H$. We denote $\langle \cdot \mid
\cdot \rangle _j$ the restriction of the scalar product (\ref{e2}) on the
lineals ${\cal L}^j$, $j=0,1$. The completion of the lineals ${\cal L}^j$
with respect to the norms generated by the appropriate scalar products
produces the Hilbert subspaces $H^j=\overline{{\cal L}}^j$. It is clear that
for the lineal ${\cal L}$ we have the orthogonal sum ${\cal L}={\cal L}%
^0\oplus {\cal L}^1$ and for the space $H$ we have the orthogonal
decomposition $H=H^0\oplus H^1$. By these means we obtain the well-known
constructions \cite{M} of the Hilbert space structure on the solutions of
the Schr\"odinger equation (\ref{e1}).

Symmetry operators for the equation (\ref{e1}) are defined as usual
as the operators which transform any solution of this equation to another
solution of the same equation. These operators realize a coordinate
representation of the Schr\"odinger algebra ${\cal G}_2$ and have the form
\begin{equation}
\label{g}
\begin{array}{c}
{\cal K}_2=-t^2\partial _t-tx\partial _x-t/2+i x^2/4,\quad {\cal K}%
_1=-t\partial _x+i x/2, \\ {\cal K}_0=i ,\quad {\cal K}_{-1}=\partial
_x,\quad {\cal K}_{-2}=\partial _t,\quad {\cal K}^0=x\partial _x+2t\partial
_t+1/2.
\end{array}
\end{equation}
The operators (\ref{g}) are defined on the dense set ${\cal L}\subset H$ and $%
{\cal L}$ is the Lie ${\cal G}_2$-module. Moreover, every operator from (\ref
{g}) is skew symmetric with respect to the scalar product (\ref{e2}) and,
consequently, with the help of the exponential mapping we may construct a
group of unitary operators \cite{Simon,M}.

Since we are not interested in the other representations of the
algebra ${\cal G}_2$ we shall denote ${\cal G}_2=\mathop{\rm span}\left\{
{\cal K}_{0,\pm 1},{\cal K}_{\pm 2},{\cal K}^0\right\} $ where $%
\mathop{\rm span}$ stands for the linear hull over the real number field $%
{\Bbb R}$ if we want to have the real algebra ${\cal G}_2({\Bbb R})$ and
over the complex number field ${\Bbb C}$ for its complex form ${\cal G}_2(%
{\Bbb C})$.

Let us pass in ${\cal G}_2({\Bbb C})$ to another basis more suitable
for our purpose. Consider $a^{\pm }=\frac 12(i {\cal K}_{-1}\mp {\cal K}_1)$%
. Operators $\left\{ a^{\pm },I\right\} $, where $I$ is the identity
operator, form the basis of the Heisenberg-Weil algebra $w_1$. Consider now
the quadratic combinations of $a^{\pm }$, i.e., $k_{\pm }=2\left( a^{\pm
}\right) ^2$, $k_0=a^{+}a^{-}+a^{-}a^{+}$. Since for every $\chi \in {\cal L}
$ equation (\ref{e1}) represents an operator identity $\partial
_x^2=-i\partial _t$, operators $k_0$ and $k_{\pm }$ acting in ${\cal L}$ may
be considered as of the first degree in $\partial _x$ and $\partial _t$. It
is easy to see that these operators form another basis in subalgebra $sl(2,%
{\Bbb C})=\mathop{\rm span}\left\{ K^0,K_{\pm 2}\right\} \subset {\cal G}_2(%
{\Bbb C})$ and ${\cal G}_2({\Bbb C})=sl(2,{\Bbb C}){%
\mathinner{\in\mkern-11mu\rule{0.15mm}
   {2.0mm}\mkern10mu}}w_1({\Bbb C})$. Its real form $sl(2,{\Bbb R})$ is
isomorphic to $su(1.1)$. Moreover, it is apparent that ${\cal L}^0$ and $%
{\cal L}^1$ are irreducible Lie $su(1.1)$-modules of the weight $1/4$ and $%
3/4$ respectively. Note that the operators $a^{\pm }$ map the lineals ${\cal %
L}^0$ and ${\cal L}^1$ one into another. We note as well the following
conjugation properties of the operators from ${\cal G}_2$ with respect to
the scalar product (\ref{e2}): $\left( a^{\pm }\right) ^{\dagger }=a^{\mp }$%
, $k_0^{\dagger }=k_0$, $\left( k_{\pm }\right) ^{\dagger }=k_{\mp }$, which
hold on the lineal ${\cal L}$.

To conclude this section we would like to notice that the procedure
specified above may be applicable to a wide class of Hamiltonians of the
form $h=-\partial _x^2+A(t)x^2+B(t)x+C(t)$. This assertion follows from the
well-known fact that the nonstationary Schr\"odinger equation with this
Hamiltonian has integrals of motion (symmetry operators) $\widetilde{a}^{\pm
}$ that depend only on the derivative $\partial _x$ and form a
representation of the Heisenberg-Weil algebra $w_1$ \cite{MalMan}. Their
quadratic combinations being apparently symmetry operators may be expressed
only through the $\partial _x$ and $\partial _t$ on the space of the
solutions of the nonstationary Schr\"odinger equation. These operators form
a representation of the algebra $su(1.1)$. The semidirect sum of
representations of $su(1.1)$ and $w_1$ gives a representation of the
Schr\"odinger algebra ${\cal G}_2$. The Hilbert space structure on the
solutions of the Schr\"odinger equation is introduced with the help of the
well-known constructions of the discrete series representation of the
Schr\"odinger algebra. The latter is built of two Lie irreducible $su(1.1)$
modules in the same way as above.

\vspace{0.5cm}

\begin{flushleft}
{\large {\bf III. HILBERT SUPERSPACE OF SQUARE INTEGRABLE FUNCTIONS} }
\end{flushleft}

\vspace{0.5cm}
{\bf A. Basic definitions}

\vspace{0.5cm} In what follows we shall use the Grassmann-valued
analysis \cite{DW,Ber,Rog,L,Vlad}. It is worth mentioning that for some
notions several distinct definitions exist in the literature. In these cases
we give the definition we use.

The more suitable approach to the superanalysis for our purpose is
the one described in Ref. \cite{Vlad} and based on the theory of functions
in Banach spaces and the theory of Banach algebras. The basis notion in this
approach is the notion of commutative Banach superalgebra introduced as
follows.

Let $\Lambda $ be a $z_2$-graded linear space $\Lambda =\Lambda
_0\oplus \Lambda _1$. When an element $a\in \Lambda _0$, it is called {\it %
even} [parity $p(a)=0$] and when $a\in \Lambda _1$ it is called {\it odd}
[parity $p(a)=1$]. The elements from $\Lambda _0$ and $\Lambda _1$ are
called {\it homogeneous}. When the structure of associative algebra with
unit $e\in \Lambda _0$ and even multiplication operation [i.e., $%
p(ab)=p(a)+p(b)$, $\mathop{\rm mod}2$ for homogeneous $a$ and $b$] is
introduced in $\Lambda $ it is called {\it superalgebra}. Superalgebra $%
\Lambda $ is called {\it commutative} if {\it supercommutator} $\left[
a,b\right] =ab-(-1)^{p(a)p(b)}ba=0$ for homogeneous $a,b\in \Lambda $.
Further, the commutative superalgebra $\Lambda =\Lambda _0\oplus \Lambda _1$
is supposed to be a Banach space with the norm $\left\| fg\right\| \le
\left\| f\right\| \cdot \left\| g\right\| $, $f,g\in \Lambda $, $\left\|
e\right\| =1$. The components $\Lambda _0$ and $\Lambda _1$ are closed
subspaces in $\Lambda $. When $\Lambda $ is defined over the real number
field ${\Bbb R}$ we obtain the real superalgebra $\Lambda ({\Bbb R})$, and
for the case of the complex number field ${\Bbb C}$ we obtain its complex
form $\Lambda ({\Bbb C})$.

Given a real superalgebra $\Lambda ({\Bbb R})$ real superspace $%
{\Bbb R}_\Lambda ^{m,n}$ of dimension $(m,n)$ over $\Lambda ({\Bbb R})$ is
defined as follows:
$$
\begin{array}{ccccccc}
{\Bbb R}_\Lambda ^{m,n} & = & \underbrace{\Lambda _0\otimes \ldots \otimes \Lambda
_0} & \otimes  & \underbrace{\Lambda _1\otimes \ldots \otimes \Lambda _1} &
= & \Lambda _0^m\otimes \Lambda _1^n. \\
&  & m &  & n &  &
\end{array}
$$
A complex superspace ${\Bbb C}_\Lambda ^{m,n}$ over $\Lambda ({\Bbb C})$ is
defined in the same way but with the help of the complex superalgebra $%
\Lambda ({\Bbb C})$. If for every point $X=(x,\theta )=(x_1,\ldots ,x_m,$ $%
\theta _1,\ldots \theta _n)\in {\Bbb R}_\Lambda ^{m,n}$ we introduce the
norm $\parallel X\parallel ^2=\parallel x\parallel ^2+\parallel \theta
\parallel ^2=\sum_{k=1}^m\parallel x_k\parallel ^2+\sum_{j=1}^n\parallel
\theta _j\parallel ^2$, then ${\Bbb R}_\Lambda ^{m,n}$ becomes a Banach
space. Every connected open set ${\Bbb O}\subset {\Bbb R}_\Lambda ^{m,n}$ is
called {\it domain} in ${\Bbb R}_\Lambda ^{m,n}$.

Let us have two superspaces ${\Bbb R}_\Lambda ^{m,n}$ and ${\Bbb R}%
_{\Lambda ^{\prime }}^{m^{\prime },n^{\prime }}$ with the norms $\parallel
\cdot \parallel $ and $\parallel \cdot \parallel ^{\prime }$, $\Lambda
\subseteq \Lambda ^{\prime }$, and a domain ${\cal O}$ in ${\Bbb R}_\Lambda
^{m,n}$. Function $f(X):{\cal O}\rightarrow {\Bbb R}_{\Lambda ^{\prime
}}^{m^{\prime },n^{\prime }}$ is called {\it continuous} in the point $X\in
{\cal O}$ if $\parallel f(X+H)-f(X)\parallel ^{\prime }\rightarrow 0$ when $%
\parallel H\parallel \rightarrow 0$. The same function is called {\it %
superdifferentiable from the left} in the point $X\in {\cal O}$ if elements $%
F_k(X)\in {\Bbb R}_{\Lambda ^{\prime }}^{m^{\prime },n^{\prime }}$, $%
k=1,\ldots ,m+n$, such that
$$
f(X+H)=f(X)+\sum_{k=1}^{m+n}H_kF_k(X)+\tau (X,H),
$$
where $\parallel \tau (X,H)\parallel ^{\prime }/\parallel H\parallel
\rightarrow 0$ when $\parallel H\parallel \rightarrow 0$ exist. The
functions $F_k(x)$ are called {\it left partial superderivatives} of $f$
with respect to $X_k$ in the point $X\in {\cal O}$:
$$
F_k(x)=\frac{\partial f(X)}{\partial X_k},\quad F_{m+j}(x)=\frac{\partial
f(X)}{\partial X_{m+j}},\quad k=1,\ldots ,m,\quad j=1,\ldots n.
$$
The expression $\sum_{k=1}^{m+n}H_k\frac{\partial f(X)}{\partial X_k}$ is
called {\it left superdifferential} of the function $f(X)$ in the point $X$.

One can find more details about superanalysis in Ref. \cite{Vlad}.

\vspace{0.5cm}
{\bf B. Hilbert superspace}

\vspace{0.5cm} Consider the real superspace ${\Bbb R}_\Lambda ^{1,1}$
defined over $\Lambda ({\Bbb R})=\Lambda _0({\Bbb R})\otimes \Lambda _1(%
{\Bbb R})$ where $\Lambda _0({\Bbb R})={\Bbb R}$ and $\Lambda _1({\Bbb R})$
has two generators $\theta $ and $\overline \theta $ with the properties $%
\theta ^2={\overline \theta}^2=\theta \overline \theta + \overline \theta
\theta =0$, $\overline {\overline \theta }= \theta$. The complex superspace $%
{\Bbb C}_\Lambda ^{1,1}$ is defined over $\Lambda ({\Bbb C})=\Lambda _0(%
{\Bbb C})\otimes \Lambda _1({\Bbb C})$ where $\Lambda _0({\Bbb C})={\Bbb C}$
and $\Lambda _1({\Bbb C})$ has the same generators $\theta $ and $\overline
\theta $.

Consider now functions from ${\Bbb R}_\Lambda ^{1,1}$ to ${\Bbb C}%
_\Lambda ^{1,1}$ of the following form: $\Psi ^0\left( t,x,\theta ,\overline{%
\theta }\right) =\psi \left( x,t\right) $, $\psi \left( x,t\right) \in H^0$
and $\Psi ^1\left( t,x,\theta ,\overline{\theta }\right) =\theta \varphi
\left( x,t\right) $, $\varphi \left( x,t\right) \in H^1$. We shall designate
the collection of the functions $\Psi ^0\left( t,x,\theta ,\overline{\theta }%
\right) $ and $\Psi ^1\left( t,x,\theta ,\overline{\theta }\right) $ as $H_{%
\overline{0}}$ and $H_{\overline{1}}$, respectively. It follows from these
constructions that $H_{\overline{0}}$ and $H_{\overline{1}}$ are linear
spaces (over the field ${\Bbb C}$), and $H_s=H_{\overline{0}}\oplus H_{%
\overline{1}}$ is a $z_2$-graded linear space of functions. The elements
from $H_{\overline{0}}$ and $H_{\overline{1}}$ are called {\it homogeneous}
with the parity $p(\Phi )=0$ when $\Phi \in H_{\overline{0}}$ and $p(\Phi )=1
$ when $\Phi \in H_{\overline{1}}$.

Define in the space $H_s$ {\it scalar product} ({\it super Hermitian
form}) as follows:
\begin{equation}
\label{ssp}\left( \Phi _1\mid \Phi _2\right) =\int \overline{\Phi }_1\left(
t,x,\theta ,\overline{\theta }\right) \Phi _2\left( t,x,\theta ,\overline{%
\theta }\right) i e^{-i \overline{\theta }\theta }dxd\theta d\overline{%
\theta }\in {\Bbb C}\,.
\end{equation}
Since the integration in superspaces is developed in Ref. \cite{Vlad} for
sufficiently smooth functions (it is a super generalization of various
integral constructions based on Riemann integral and not on Lebesgue
integral) we should make more precise the sense of integral in (\ref{ssp}).
If functions $\Phi _1$ and $\Phi _2$ are defined by their homogeneous
components $\Phi _l\left( x,\theta ,\overline{\theta }\right) =\Phi
_l^0\left( x,\theta ,\overline{\theta }\right) +\Phi _l^1\left( x,\theta ,%
\overline{\theta }\right) $, $\Phi _l^0\left( x,\theta ,\overline{\theta }%
\right) =\chi _l^0\left( x\right) \in H_{\overline{0}}$, and $\Phi
_l^1\left( x,\theta ,\overline{\theta }\right) =\theta \chi _l^1\left(
x\right) \in H_{\overline{1}}$, $l=1,2$, and functions $\chi _l^j(x)$, $j=0,1
$, are sufficiently smooth, then we may interpret the integral (\ref{ssp})
in the sense defined in Ref. \cite{Vlad}. In our case this integral becomes
equal to a product of two integrals. The first one is a conventional
integral with respect to the variable $x$ and the second one is an integral
with respect to the Grassmann variables $\theta $ and $\overline{\theta }$.
The only integral with respect to the Grassmann variables different from
zero is $\int \overline{\theta }\theta d\theta d\overline{\theta }=1$. Thus,
for the integral (\ref{ssp}) we obtain the expression
\begin{equation}
\label{ssp1}\left( \Phi _1\mid \Phi _2\right) =\left( \Phi _1^0\mid \Phi
_2^0\right) _0+\left( \Phi _1^1\mid \Phi _2^1\right) _1,
\end{equation}
$$
\left( \Phi _1^j\mid \Phi _2^j\right) _j=i ^j\langle \chi _1^j\mid \chi
_2^j\rangle _j,~\chi _l^j\in H^j,~l=1,2,~j=0,1.
$$
We note that the spaces $H_{\overline{0}}$ and $H_{\overline{1}}$ are
mutually orthogonal with respect to the scalar product (\ref{ssp}) and are
complete in the sense we shall make more precise so that $(\cdot |\cdot )_j$%
, $j=0,1$, are the restrictions of the scalar product (\ref{ssp}) on the
spaces $H_{\overline{j}}$.

In the case when functions $\chi _l^j\in L_2({\Bbb R})$ are not
sufficiently smooth for applying the definition of the integral given in
Ref. \cite{Vlad}, we directly apply the formula (\ref{ssp1}) for calculating
the integral (\ref{ssp}). We remind the reader that the scalar product $%
\langle \cdot \mid \cdot \rangle $ in $L_2({\Bbb R})$ is defined with the
help of the Lebesgue integral. We will notice that the formula (\ref{ssp1})
is in accord with the definition of the super-Hermitian form in the abstract
Hilbert superspace given in Ref. \cite{GN}. The super-Hermitian form (\ref
{ssp1}) is positive definite in the sense that the Hermitian forms $\langle
\cdot \mid \cdot \rangle _j$, $j=0,1$, from which it is expressed are
positive definite.

The super-Hermitian form generates a norm in $H_s$. For every $\Phi
=\Phi ^0+\Phi ^1\in H_s$, $\Phi ^0=\chi ^0(x,t)$, and $\Phi ^1=\theta \chi
^1(x,t)$ we put by definition
\begin{equation}
\label{e5}\parallel \Phi \parallel ^2\equiv \left| \left( \Phi \mid \Phi
\right) \right|^2 =\parallel \chi ^0\parallel _0^2+\parallel \chi ^1\parallel
_1^2,
\end{equation}
where $\parallel \cdot \parallel _j$ are the norms in $H^j$, $j=0,1$,
generated by the appropriate scalar products. It is not difficult to see
that the properties of the norm so defined correspond to the axioms of the
conventional norm: (i) $\parallel \Phi \parallel \ge 0$, (ii) $\parallel
\Phi \parallel =0$ if and only if $\Phi =0$, (iii) $\parallel c\Phi
\parallel =\mid c\mid \cdot \parallel \Phi \parallel $, $\forall c\in {\Bbb C%
}$, (iv) $\parallel \Phi _1+\Phi _2\parallel \le \parallel \Phi _1\parallel
+\parallel \Phi _2\parallel $. It follows that $H_s$ is a normed space in
the usual sense. Conditions (i), (iii), and (iv) mean that the norm is a
convex functional in $H_s$ (see, e.g., Ref. \cite{Naim}). Condition (ii)
means that the set $\left\{ \parallel \cdot \parallel \right\} $ formed from
a single convex functional is sufficient for defining a (strong) topology in
$H_s$. The space $H_s$ becomes a locally convex topological space \cite{Naim}%
. Just in this sense we shall understand the completeness of the space $H_s$
which we shall call the {\it Hilbert superspace}. In fact, this signifies
that the space $H_s$ contains only linear functions of the variable $\theta $
with the coefficients from $H$. Since the functions $\Psi _n^0\left(
t,x,\theta ,\overline{\theta }\right) =\psi _n\left( x,t\right) $ and $\Psi
_n^1\left( t,x,\theta ,\overline{\theta }\right) =\theta \varphi _n\left(
x,t\right) $ form bases in the spaces $H_{\overline{0}}$ and $H_{\overline{1}%
}$, respectively, we have obtained a {\it separable} Hilbert superspace. It
is worth noticing that other definitions of the Hilbert superspace exist
\cite{HSS}.

\hspace{0.5cm}

\begin{flushleft}
{\large {\bf IV. ATYPICAL COORDINATE REPRESENTATION OF $osp(2/2)$} }
\end{flushleft}

\vspace{0.5cm}
We may now define the action of operators in the
space $H_s$. Let us put $K_0=k_0=h/2$ and $K_{\pm }=k_{\pm }$. These
operators by definition act only on the variable $x$ and do not affect the
Grassmann variable $\theta $. This signifies that they have the even parity.
Since the operators $k_0$ and $k_{\pm }$ are defined on the lineal ${\cal L}=%
{\cal L}^0\oplus {\cal L}^1$, the operators $K_0$ and $K_{\pm }$ are defined
on the lineal ${\cal L}_s={\cal L}_{\overline{0}}\oplus {\cal L}_{\overline{1%
}}$ over the field ${\Bbb C}$, where ${\cal L}_{\overline{0}}$ is the lineal
over ${\Bbb C}$ of the even functions $\Psi _n^0(t,x,\theta ,\overline{%
\theta })$ and ${\cal L}^1$ is the lineal over ${\Bbb C}$ of the odd ones $%
\Psi _n^1(t,x,\theta ,\overline{\theta })$. It is clear that closure $%
\overline{{\cal L}}_{\overline{0}}$ of the lineal ${\cal L}_{\overline{0}}$
with respect to the norm (\ref{e5}) gives $H_{\overline{0}}$ and similar
closure $\overline{{\cal L}}_{\overline{1}}$ gives $H_{\overline{1}}$.
Moreover, $\overline{{\cal L}}_s=\overline{{\cal L}}_{\overline{0}}\oplus
\overline{{\cal L}}_{\overline{1}}=H_s$. Operators $K_0$ and $K_{\pm }$ form
a basis of subalgebra $su(1.1)$ of the superalgebra under construction and
lineals ${\cal L}_{\overline{0}}$ and ${\cal L}_{\overline{1}}$ are Lie
irreducible $su(1.1)$-modules.

We may define operators of the left multiplication by the variable $%
\theta $ and the left differentiation $\partial _\theta =\overrightarrow{{%
\textstyle {\frac{\partial }{\partial \theta }}}}$ (we define the left action
of the operators on vectors) for the elements from ${\cal L}_s$. It is clear
that in our case $\forall \Phi \in {\cal L}_s$ we have $\theta \Phi \notin
{\cal L}_s$ and $\partial _{\theta }\Phi \notin {\cal L}_s$. Nevertheless,
the same operators may be defined not only for the elements from ${\cal L}_s$
but for every linear function of $\theta $. Therefore operator $B=\frac
14\left(\theta \partial _\theta -\partial _\theta \theta \right) $ is
defined on the functions from ${\cal L}_s$ and ${\cal L}_s$ is its invariant
space. Moreover, it is easy to see that $B\Phi =-\frac 14 (-1)^{p(\Phi
)}\Phi $ for every homogeneous $\Phi \in {\cal L}_s$.

Operators $a^{\pm }$ map the lineals ${\cal L}^0$ and ${\cal L}^1$
one into another. With their help we construct the odd sector of the
superalgebra under construction: $V_{\pm }=\sqrt{2}a^{\pm }\theta $ and $%
W_{\pm }=\sqrt{2}a^{\pm }\partial _\theta $. Operators $V_{\pm }$ map
vectors from ${\cal L}_{\overline{0}}$ to vectors from ${\cal L}_{\overline{1%
}}$. Operators $W_{\pm }$ realize the inverse mapping. In addition, $V_{\pm
}\Phi ^1=0$ $\forall \Phi ^1\in {\cal L}_{\overline{1}}$ and $W_{\pm }\Phi
^0=0$ $\forall \Phi ^0\in {\cal L}_{\overline{0}}$.

It is an easy exercise to check that the set of operators $\left\{
K_0,K_{\pm },B, V_{\pm },W_{\pm }\right\} $ is closed with respect to the
supercommutator $\left[ A,C\right] =AC-\left( -1\right) ^{p(A)p(B)}CA$ and
generalized Jacobi identity holds. The nonzero supercommutators are written
as follows:%
$$
\left[ K_0,K_{\pm }\right] =\pm K_{\pm },\,\,\left[ K_{-},K_{+}\right]
=2K_0,\,\,\left[ K_0,V_{\pm }\right] =\pm \frac 12V_{\pm },\,\,\left[
K_0,W_{\pm }\right] =\pm \frac 12W_{\pm },
$$
$$
\left[ K_{\pm },V_{\mp }\right] =\mp V_{\pm },\,\,\left[ K_{\pm },W_{\mp
}\right] =\mp W_{\pm },\,\,\left[ B,V_{\pm }\right] =\frac 12V_{\pm
},\,\,\left[ B,W_{\pm }\right] =-\frac 12W_{\pm },
$$
$$
\left[ V_{\pm },W_{\pm }\right] =K_{\pm },\,\,\left[ V_{\pm },W_{\mp
}\right] =K_0\mp B.
$$

Vector $\Psi _0^0$ has the properties%
$$
K_0\Psi _0^0=\frac 14\Psi _0^0,\,\,B\Psi _0^0=-\frac 14\Psi
_0^0,\,\,K_{-}\Psi _0^0=V_{-}\Psi _0^0=W_{\pm }\Psi _0^0=0.
$$
It follows that the set of operators $\left\{ K_0,K_{\pm },B,V_{\pm },W_{\pm
}\right\} $ realizes an atypical (coordinate) representation of the abstract
orthosymplectic superalgebra $osp(2/2)=osp(2/2)_{\overline{0}}\oplus
osp(2/2)_{\overline{1}}$, where $osp(2/2)_{\overline{0}}=\mathop{\rm span}%
\left\{ K_0,K_{\pm },B\right\} $ and $osp(2/2)_{\overline{1}}=%
\mathop{\rm span}$ $\left\{ V_{\pm },W_{\pm }\right\} $.

The nonrelativistic free particle Hamiltonian $h=-\partial
_x^2=\left( a^{+}+a^{-}\right) ^2=\frac 12K_{+}+\frac 12K_{-}+K_0$ is an
element of $osp(2/2)$ superalgebra and, consequently, this algebra is a
dynamical supersymmetry algebra for this system. Moreover, its
representation space ${\cal L}_s$ is the space of solutions of the free
particle Schr\"odinger equation
$$
i \partial _t\Psi \left( t,x,\theta ,\overline{\theta }\right) =h\Psi
\left( t,x,\theta ,\overline{\theta }\right) .
$$

Given the super-Hermitian forms (\ref{ssp}) and (\ref{ssp1}) w%
e define operator $A^{+}$ {\it superadjoint} to an $A$%
. An operator $A\in osp(2/2)$ is defined on the dense set ${\cal L}_s$ in $%
H_s$. Then, for every homogeneous element $A\in osp(2/2)$, element $\Phi
_1^{*}\in H_s$ is uniquely defined by the equation
\begin{equation}
\label{AS}\left( \Phi _1^{*}|\Phi _2\right) =(-1)^{p(\Phi _1)p(A)}\left(
\Phi _1|A\Phi _2\right) ,\quad \Phi _2\in {\cal {L}}_s,
\end{equation}
for a homogeneous $\Phi _1\in H_s$. Therefore we may put $\Phi
_1^{*}=A^{+}\Phi _1$. The domain of definition of operator $A^{+}$ is the
collection of all $\Phi _1\in H_s$ which verify the equation (\ref{AS}).

The Hilbert superspace introduced here is quite analogous to the
conventional Hilbert space and we may use many conventional definitions
(see, e.g., Refs. \cite{Neum,Simon,Naim}). In particular, the definition of
a closed operator remains unchanged. Then, since the operator $h$ is
essentially self-adjoint in $H$, the operator $K_0$ is essentially
self-adjoint in $H_s$. Operator $B$ is restricted and consequently closed in
$H_s$. It is not difficult to see that $B^{+}=B$. The operators $a$ and $%
a^{+}$ are defined in ${\cal L}\subset H$ and are mutually conjugated. This
involves the following conjugation properties $K_0^{+}=K_0$, $K_{\pm
}^{+}=K_{\mp }$, $B^{+}=B$, $V_{\pm }^{+}=iW_{\mp }$and $W_{\pm
}^{+}=iV_{\mp }$, valid in ${\cal L}_s$. Moreover, $\forall A\in osp(2/2)$
the following relations hold in ${\cal L}_s$: $(A^{+})^{+}=A$, $%
(AC)^{+}=(-1)^{p(A)p(C)}C^{+}A^{+}$, and  $\left[ A,C\right] ^{+}=-\left[
A^{+},C^{+}\right] $.

Given a $z_2$-graded linear space $L_s=L_{\overline{0}}\oplus L_{%
\overline{1}}$ [for example, the lineal ${\cal L}_s$ or the superalgebra $%
osp(2/2)$] we should have the possibility to define the multiplication of
the elements from $L_s$ on the elements from a complex commutative Banach
superalgebra $\Lambda ({\Bbb C})=\Lambda _0({\Bbb C})\oplus \Lambda _1({\Bbb %
C})$. In particular, we need definition of the $\Lambda ({\Bbb C})$ envelope
of the second kind $\widetilde{L}_s$ of the space $L_s$. This definition is
similar to the definition of the Grassmann envelope of the second kind of
the space $L_s$ ( Ref. \cite{Ber}), where the role of a Grassmann algebra
plays the algebra $\Lambda ({\Bbb C})$: $\widetilde{L}_s=(\Lambda _0({\Bbb C}%
)\otimes L_{\overline{0}})\oplus (\Lambda _1({\Bbb C})\otimes L_{\overline{1}%
})=(\Lambda ({\Bbb C})\otimes L_s)_{\overline{0}}$. The elements from $%
\Lambda ({\Bbb C})$ play the role of supernumbers, the elements from $%
\Lambda _0({\Bbb C})$ play the role of $c$-numbers, and the elements from $%
\Lambda _1({\Bbb C})$ play the role of $a$-numbers. This terminology
corresponds to Ref. \cite{DW}.

Let us make more precise the definition of the complex conjugation
in $\widetilde{L}_s$ and $\Lambda ({\Bbb C})$. Put by definition
\begin{equation}
\label{e6}\overline{\beta _1\beta _2}=\overline{\beta }_1\overline{\beta }%
_2,\quad \overline{\beta \Phi }=\overline{\beta }\,\overline{\Phi },\quad
\forall \Phi \in L_s,\quad \forall \beta ,\beta _1,\beta _2\in \Lambda (%
{\Bbb C}).
\end{equation}
This definition differs from a one widely used in literature $\overline{%
\beta _1\beta _2}=\overline{\beta }_2\overline{\beta }_1$, $\beta _1,\beta
_2\in \Lambda ({\Bbb C})$ (Refs. \cite{Ber,DW}) which gives for the product $%
\beta \overline{\beta }$ a real value independently on the parity of $\beta $%
. We shall use the definition (\ref{e6}) since in the other case one faces
some inconsistencies. In particular, the super K\"ahler two-form becomes
neither real nor imaginary, and it is difficult to establish the
correspondence between physical observables and self-superadjoint operators
\cite{GN}.

We shall use the expression (\ref{e5}) for calculating of the scalar
product of the elements from $\widetilde H_s$. Definition (\ref{ssp})
realizes in this case the following mapping: $\widetilde{H}_s\otimes
\widetilde{H}_s\rightarrow \Lambda _0({\Bbb C}) $. The rule of manipulation
with the supernumbers in the scalar product
$$
\left( \beta _1\Phi _1\mid \beta _2\Phi _2\right) =\left( -1\right)
^{p\left( \Phi _1\right) p\left( \beta _2\right) }\overline{\beta }_1\beta
_2\left( \Phi _1\mid \Phi _2\right) ,
$$
where $\Phi _1$ and $\beta _2$ are homogeneous elements from $H_s$ and $%
\Lambda ({\Bbb C})$ and the rule of complex conjugation for the scalar
product of homogeneous elements
$$
\overline{\left( \Phi _1\mid \Phi _2\right) }=\left( -1\right) ^{p\left(
\Phi _1\right) p\left( \Phi _2\right) }\left( \Phi _2\mid \Phi _1\right) .%
$$
follow from Eq. (\ref{ssp}) as well.

\vspace{0.5cm}
\newpage
\begin{flushleft}
{\large {\bf V. FREE PARTICLE SUPERCOHERENT STATES} }
\end{flushleft}

\vspace{0.5cm}
Supercoherent states are the direct generalization
\cite{FKNT} of coherent states for the conventional (non-super) Lie groups
and algebras \cite{P}.

We follow the definition of supergroup given in Ref. \cite{Rog}. An $%
(m,n)$-dimensional supergroup $G$ is both an abstract group and an $(m,n)$%
-dimensional superanalytic supermanifold $S_\Lambda ^{m,n}$ with
superanalytic mapping $G\otimes G\rightarrow G:(g_1,g_2)\rightarrow
g_1g_2^{-1}$. Superanalytic supermanifold $S_\Lambda ^{m,n}$ is defined as a
Hausdorf space with an atlas such that $S_\Lambda ^{m,n}$ is locally
homeomorphic to a flat superspace $R_\Lambda ^{m,n}$ and the transition
functions are superanalytic.

An operator of left translations on a supergroup has been used in
Refs. \cite{FKNT,G} for constructing supercoherent states for the algebra $%
osp(1/2)$ and in Ref. \cite{GN} for the algebra $osp(2/2)$. Using the same
approach we pass in $osp(2/2,{\Bbb C})$ to {\it super Hermitian base,}
$$
X_1=K_0,\,X_2=B,\,X_3=K_{+}+K_{-},\,X_4=i (K_{+}-K_{-}),
$$
$$
X_5=V_{+}-i W_{-},\,X_6=V_{-}-i W_{+},\,X_7=W_{+}-i V_{-},\,X_8=W_{-}-i %
V_{+},
$$
which has the property $X_j^{+}=(-1)^{p(X_j)}X_j$. Further, the Grassmann
envelop $\widetilde{osp}(2/2)$ of the algebra ${osp}(2/2)$ over the real
Grassmann algebra $G(2)=G^0(2)\oplus G^1(2)$ should be considered. An
arbitrary element $\widetilde{X}$ from $\widetilde{osp}(2/2)$ has the form
\begin{equation}
\label{Tg}\widetilde{X}=\sum_{j=1}^4\xi _j^0X_j+\sum_{j=1}^4\xi
_j^1X_{4+j},\,\,\xi _j^0\in G^0(2),\,\,\xi _j^1\in G^1(2).
\end{equation}
The set of left translations on the supergroup $OSp(2/2)$ is defined as
follows \cite{FKNT}: $T(g)=\exp (i\widetilde{X})$, $g\in OSp(2/2)$.

The highest symmetry vector (fiducial state) in $H_s$ is $\Psi _0^0$%
. Its isotropy subalgebra consists of Cartan subalgebra of $osp(2/2,{\Bbb C})
$ which is spanned of $\left\{ B,K_0\right\} $, all lowering operators $%
\left\{ K_{-},V_{-},W_{-}\right\} $, and one raising operator $W_{+}$. The
latter is a consequence of the fact that we have obtained the atypical
representation of the $osp(2/2)$ superalgebra \cite{GN}. Therefore, the $%
osp(2/2)$ coherent states in this case are the $osp(1/2)$ coherent states as
well and are labeled by one complex parameter $z\in {\Bbb C}$, $\left|
z\right| <$ , and one Grassmann parameter $\alpha $. Since $z$ parametrizes
the unit disc ${\cal D}^{(1)}$, the corresponding supermanifold, realized in
terms of coordinates $\left( z,\alpha \right) $, is called $N=1$ superunit
disc and denoted by ${\cal D}^{(1|1)}\equiv OSp(2/2)/U(1/1)$ where subgroup $%
U(1/1)$ has the generators $K_0$, $B$, and $W_{\pm }$. This reasoning leads
to the following translation operator suitable in our case:
\begin{equation}
\label{D}D^{\prime }(z,\alpha )=\exp \left( zK_{+}-\overline{z}K_{-}+\alpha
V_{+}-i \overline{\alpha }W_{-}\right) ,\,\,|z|<1.
\end{equation}
where the use of the complex variables $(z,\alpha )$ instead of the real
supernumbers $\xi _j^l$ has been made.

The action of operator (\ref{D}) is defined on the elements from $%
{\cal L}_s\subset H_s$. It maps the lineal ${\cal L}_s$ onto ${\widetilde{H}}%
_s$ defined over $\Lambda ({\Bbb C})=\Lambda _0({\Bbb C})\oplus \Lambda _1(%
{\Bbb C})$, where $\Lambda _0({\Bbb C})={\Bbb C}$ and $\Lambda _1({\Bbb C})$
is defined over ${\Bbb C}$ with the help of two Grassmann generators $\xi $,
$\overline{\xi }$ and $\alpha ,\overline{\alpha }\in \Lambda _1({\Bbb C})$.
Moreover, this operator preserves the value of the scalar product $%
(D^{\prime }(z,\alpha )\Phi _1\mid D^{\prime }(z,\alpha )\Phi _2)=(\Phi
_1\mid \Phi _2)$ $\forall \Phi _{1,2}\in {\cal L}_s$. This property
characterizes a {\it superisometric} operator.

To rewrite operator (\ref{D}) in the form of the ordered exponential
factors we may use superextension \cite{SBCH} of the well-known
Baker-Campbell-Hausdorf \cite{BCH} relation. However, the existence of the
relations $K_{-}\Psi _0^0=0$ and $W_{-}\Psi _0^0=0$ make it possible to use
a simpler translation operator $D(z,\alpha )=\exp \left( zK_{+}+\alpha
V_{+}\right) $, instead. This operator does not preserve the scalar product.
Therefore, we have to introduce a normalizing constant. Thus, for
supercoherent states we obtain the following relation:
\begin{equation}
\label{CS}
\begin{array}{c}
\Psi _{z\alpha }(t,x,\theta ,
\overline{\theta })=N^{\prime }\exp \left( zK_{+}+\alpha V_{+}\right) \Psi
_0^0(t,x,\theta ,\overline{\theta }) \\ =N\left( \psi _z(x,t)+\sqrt{2}\alpha
\theta \varphi _z(x,t)\right)
\end{array}
\end{equation}
where
$$
\psi _z(x,t)=\left( \frac{\sigma +\overline{\sigma }}{4\pi }\right)
^{1/4}\left( \sigma +i t\right) ^{-1/2}\exp \left[ \frac{-x^2}{4(\sigma +i %
t)}\right] ,
$$
$$
\varphi _z(x,t)=a^{+}\psi _z(x,t)=-\frac{i x}4\frac{1+\sigma }{\sigma +i t}%
\psi _z(x,t),
$$
$$
\sigma =\frac{1-z}{1+z},\quad N=1+\frac{i \overline{\alpha }\alpha }{4(1-z%
\overline{z})},\quad \left| z\right| <1.
$$
Function $\psi _z(x,t)$ is the free particle coherent state obtained by
applying of the displacement operator for the algebra $su(1.1)$ to the
lowest vector $\psi _0(x,t)$ of the representation with the weight $k^0=1/4$%
, and $\varphi _z(x,t)$ is an analogous one (but non normalized to unity)
corresponding to the weight $k^1=3/4$.

\vspace{0.5cm}

\begin{flushleft}
{\large {\bf VI. DISCUSSION AND CONCLUDING REMARKS} }
\end{flushleft}

\vspace{0.5cm}
The coherent states of the abstract orthosymplectic
superalgebra $osp(2/2)$ are studied in detail in Ref. \cite{GN}.. If we
expand the functions $\psi _z(x,t)$ and $\varphi _z(x,t)$ in terms of the
basis functions $\psi _n(x,t)$ and $\varphi _n(x,t),$
$$
\psi _z(x,t)=(1-z\overline{z})^{1/4}\sum\limits_{n=0}^\infty z^n\sqrt{\frac{%
\Gamma (n+\frac 12)}{n!\Gamma (\frac 12)}}\psi _n(x,t),
$$
$$
\varphi _z(x,t)=\frac 12\left( 1-z\overline{z}\right)
^{1/4}\sum\limits_{n=0}^\infty z^n\sqrt{\frac{\Gamma (n+\frac 32)}{n!\Gamma
(\frac 32)}}\varphi _n(x,t),
$$
we obtain the same formula that those given in Ref. \cite{GN} for the
atypical abstract $OSp(2/2)$ coherent states at $\tau =1/4$ and $b=-1/4$. In
that paper the geometric properties of the coherent states supermanifold are
studied. It is established that their underlying geometries turn out to be
those of supersymplectic $OSp(2/2)$ homogeneous space possessing the
super-K\"ahler structure; superunitary irreducible representation of $%
OSp(2/2)$ supergroup in the super-Hilbert space of the superholomorphic in
the superunit disc ${\cal D}^{(1|2)}$ (for the atypical in ${\cal D}^{(1|1)}$%
) functions is explicitly constructed.

Given the supercoherent states (\ref{CS}) and the scalar product (%
\ref{ssp}) we can calculate the classical observables in phase space ${\cal D%
}^{(1|1)}$. These are the covariant Berezin symbols of the $osp(2/2)$ super
algebra generators: $H^{cl}=\langle \overline{z}\overline{\alpha }|H|%
\overline{z}\overline{\alpha }\rangle $, $H\in osp(2/2)$. Our calculation
gives the following result:
$$
K_0^{cl}=\frac 14\frac{1+\left| z\right| ^2}{1-\left| z\right| ^2}K_\alpha
,\quad K_{+}^{cl}=\frac z{2\left( 1-\left| z\right| ^2\right) }K_\alpha
,\quad K_{-}^{cl}=\frac{\overline{z}}{2\left( 1-\left| z\right| ^2\right) }%
K_\alpha ,
$$
where
$$
K_\alpha =\left( 1+\frac{i \overline{\alpha }\alpha }{1-\left| z\right| ^2}%
\right)
$$
and
\begin{equation}
\label{V}V_{+}^{cl}=\frac{i \alpha }{2\left( 1-\left| z\right| ^2\right) }%
,\quad V_{-}^{cl}=\frac{i \alpha \overline{z}}{2\left( 1-\left| z\right|
^2\right) },
\end{equation}
$$
W_{+}^{cl}=\frac{-\overline{\alpha }z}{2\left( 1-\left| z\right| ^2\right) }%
,\quad W_{-}^{cl}=\frac{-\overline{\alpha }}{2\left( 1-\left| z\right|
^2\right) }.
$$
Note that the even quantities $K_0^{cl}$ and $K_{\pm }^{cl}$ completely
coincide with those given in Ref. \cite{GN} at $\tau =1/4$ and $b=-1/4$, but
for the odd ones we have the different sign. This difference is due to the
phase factor $(-i )^n$ in the basis functions (\ref{bf}).

Using the potential of the super-K\"ahler metric $f(z,\overline{z}%
,\alpha ,\overline{\alpha })=\log \left| \langle 0|\overline{z}\overline{%
\alpha }\rangle \right| ^{-2}$, we may calculate the supersimplectic form $%
\omega $ and then the Hamiltonian vector superfields $X_H$ associated to a
classical observables $H^{cl}$. The same supersymplectic form $\omega $ is
used to define a Poisson superbracket in the space of smooth functions on $%
{\cal D}^{(1|1)}$ and obtain by these means a Poisson superalgebra. All
these quantities are the straightforward generalization of the usual
(nonsuper) Hamiltonian mechanics (see, e.g., Ref. \cite{Arnold}), which in
our case is the Hamiltonian mechanics of the free particle in ${\cal D}%
^{(1|1)}$ phase superspace. The geometric quantization of this classical
mechanics gives the quantum mechanics of the free particle we started from,
but in the superholomorphic representation. The reader may find the detailed
calculations in Ref. \cite{GN}.

We will now discuss another interpretation of our results which is a
generalization of the conventional (nonsuper) interpretation of the free
particle squeezed states presented in Ref. \cite{NT}. Note that since $\psi
_z(x)$ is an even function and $\varphi _z(x)$ is an odd one, we have $%
\langle \psi _z\left| x\right| \psi _z\rangle =\langle \varphi _z\left|
x\right| \varphi _z\rangle =0$ and $\langle \psi _z\left| p\right| \psi
_z\rangle =\langle \varphi _z\left| p\right| \varphi _z\rangle =0$ where $p=-%
i \partial /\partial x$. Using the expressions of $x$ and $p$ in terms of
the operators $a^{\pm }$: $x=2pt+2i \left( a^{+}-a^{-}\right) $ and $%
p=-\left( a^{+}+a^{-}\right) $, we express the products $x\theta $ and $%
p\theta $ in terms of the superalgebra generators $V_{\pm }$:%
$$
p\theta =-\frac 1{\sqrt{2}}\left( V_{+}+V_{-}\right) ,\quad x\theta
=2tp\theta +i \sqrt{2}\left( V_{+}-V_{-}\right) .
$$
With the help of the expressions for $V_{\pm }^{cl}$ (\ref{V}) we find the
expectation values of these quantities in the state $\Psi _{z\alpha }:$%
$$
\langle p\theta \rangle _{z\alpha }=p_0\overline{\alpha },\quad \langle
x\theta \rangle _{z\alpha }=\left( 2p_0t+x_0\right) \overline{\alpha },
$$
where%
$$
x_0=-\frac{1-z}{\sqrt{2}\left( 1-z\overline{z}\right) },\quad p_0=-\frac{i %
\left( 1+z\right) }{2\sqrt{2}\left( 1-z\overline{z}\right) }.
$$
If now we pass from the variables $z$ and $\overline{z}$ to $p_0$ and $x_0$
by putting $z=\left( i p_0+\frac 12x_0\right) /\left( i p_0\right. $ $%
-\left. \frac 12x_0\right) $, we may conclude that the trajectory of a
particle becomes a straight line in the odd sector of the superspace whereas
in the even sector the particle is immovable because of the conditions $%
\langle \Psi _{z\alpha }\left| x\right| \Psi _{z\alpha }\rangle =0$ and $%
\langle \Psi _{z\alpha }\left| p\right| \Psi _{z\alpha }\rangle =0$.

In this paper a simpler example of the space of square integrable
superfunctions is given. This space may be considered as a realization of a
Hilbert super space. It is clear that in more complex cases we need to have
a theory of measure for superspaces. In particular, to give a mathematically
rigorous general concept of square integrable superfunctions the super
generalization of the Lebesgue integral based on the Lebesgue measure is
indispensable. We now have many interesting results obtained in
supersymmetric quantum mechanics \cite{Coop} but a mathematically rigorous
and consistent base of this theory is far from completion.

As a final comment we note that our constructions of the Hilbert
superspace $H_s$ are based on a natural grading of the conventional Hilbert
space $H=H^0\oplus H^1$. Therefore, these constructions are applicable not
only to the free particle but to every system for which such a decomposition
exists. In particular, minor modifications are necessary for obtaining a
Hilbert superspace structure on the solutions of the Schr\"odinger equation
with a Hamiltonian quadratic in $x$. Further, the representation of the $%
osp(2/2)$ superalgebra obtained in this paper is based on an infinite
dimensional representation of the Schr\"odinger algebra ${\cal G}_2$. It
follows that every quantum system with the same symmetry algebra may be
treated as a system possessing a dynamical $osp(2/2)$ supersymmetry. With
the help of the operators $\widetilde{a}$ and $\widetilde{a}^{+}$ a
representation of $osp(2/2)$ superalgebra sutable for this case may be
constructed. Exponential mapping of the $OSp(2/2)$ generators gives
superisometric supergroup operator translation which produces the
supercoherent states of the system under consideration in the same way as it
was made above.

\vspace{0.5cm}
\begin{flushleft} \large {\bf ACKNOWLEDGMENT}
\end{flushleft}
This research has been partially supported by RFBR grant No 97-02-16279.

\end{document}